\documentclass[12pt]{article}

\input{tcilatex}

\begin{document}

\begin{titlepage}
\setcounter{page}{1}
\title{On the (im)possibility of a supersymmetric extension of NGT}
\author{Karim Ait Moussa\thanks{E-mail : karim.aitmoussa@wissal.dz} \\
\small Laboratoire de Physique Th\'eorique \\
\small D\'epartement de Physique, Facult\'e des Sciences, \\
\small Universit\'e Mentouri, Constantine 25000, Algeria.}
\bigskip
\date{October 10, 2002}
\maketitle
\begin{abstract}
We investigate the possibility of constructing a locally
supersymmetric extension of NGT (Nonsymmetric Gravitation Theory),
based on the graded extension of the Poincar\'e group. In the
framework of the simple model that we propose, we end up with a no-go
result, namely the impossibility of cancelling some linear
contribution in the  gravitino field. This drawback seems to seriously
undermine the construction of a supergravity based on NGT.
\end{abstract}
\end{titlepage}

\setcounter{page}{2}

\section{Introduction}

Nonsymmetric gravitation theory (NGT) is a theory of gravitation based on a
nonsymmetric metric and affine connection, first proposed by Moffat \cite{1}
. It stemmed from the Einstein unified field theory\cite{2}, the aim of
which was to unify the gravitational and electromagnetic interactions in a
geometric framework, by introducing a nonsymmetric ''metric'' tensor $g_{\mu
\nu }=g_{\left( \mu \nu \right) }+g_{\left[ \mu \nu \right] }$, with the
hope of relating the antisymmetric part $g_{\left[ \mu \nu \right] }$ to the
Maxwell field, the symmetric part describing gravitation. In contrast, the
nonsymmetric metric $g_{\mu \nu }$ \emph{fully }describes gravitation in
NGT. A considerable amount of work has been devoted to the investigation of
the phenomenological consequences of this theory \cite{3}, as well as to its
geometrical interpretation \cite{4,5,6}. The early version of this theory
describes, in the linear approximation \cite{7}, the propagation of a
massless spin 2 graviton and a scalar particle, the so called skewon, which
would be the exchange particle of a long range fifth force, in addition to
the gravitational one. Damour and \emph{al.} \cite{8} have assailed NGT,
claiming it to be theoretically inconsistent, especially regarding the
expansion of the antisymmetric part of the metric $g_{\left[ \mu \nu \right]
}$ about an Einstein (symmetric) background, and suggested the addition of a
cosmological term to cure this pathological behaviour, thereby giving a mass
to the skewon. This issue have more recently been touched upon by Bekaert
and \emph{al. }\cite{9}, in the context of coupling gravity to antisymmetric
gauge fields. Later on, after having responded to the criticisms of Damour
and \emph{al. }in a series of papers \cite{10}, a new consistent version
(called massive NGT) was given by Moffat \cite{11}; it differs from the
earlier massless one by additional non-derivative terms in the action that
defines the theory (more details will be given below). Besides the massless
graviton, the resulting theory describes the propagation of a massive spin
one particle associated with the antisymmetric part of the metric\cite{12}.

Our aim in this work is to investigate the possibility of building a
supergravity based on NGT, that is a locally supersymmetric field theory
having NGT as its bosonic sector (or part of it). Ordinary supergravity \cite
{13} is the gauge theory of the supersymmetric extension of the Poincar\'{e}
group \cite{14}, the local invariance group of general relativity. In this
connection, one should use the supersymmetric extension of the local group
of invariance of NGT, that is $U\left( 3,1,H\right) $, the (pseudo)unitary
group of matrices with elements in the ring of hyperbolic complex numbers $H$
\cite{5,6}. These numbers are introduced owing to the isomorphism of $%
U\left( 3,1,H\right) $ and $GL(4,{R)}$ \cite{5}. However, it is well known
that the spinorial representations (inherent in a supersymmetric extension)
of $GL(4,{R)}$ are infinite dimensional \cite{15}. To circumvent this
technical (as well as physical, in the interpretation in terms of particles)
difficulty, one can take advantage of the fact that the homogeneous Lorentz
group is the real subgroup of $U\left( 3,1,H\right) $. Thus, in a first
trial, we restrict ourselves to the investigation of the possibility of the
construction of a \emph{minimal} extension of NGT invariant under the
supersymmetric extension of the Poincar\'{e} group. By minimal, we mean the
achievement of invariance without additional ''kinetic'' (that is containing
derivatives) terms to the NGT Lagrangian.

In section 2, we briefly review the field structure of NGT, with emphasis on
the role of the group $U\left( 3,1,H\right) $ in the anholonomic (or
vierbein) formulation \cite{6} that is relevant to our work. In section 3,
we set out our model by introducing the different fields that enter it and
their Lagrangians. In section 4, we put forward the arguments that will lead
us to our no-go result, namely the impossibility of cancelling the
supersymmetric variation of the ''kinetic'' Lagrangian of NGT, without the
addition to this Lagrangian of terms containing derivatives. We draw our
conclusions in the last section.

\section{Field structure and anholonomic formulation of NGT}

The nonsymmetric gravitation theory \cite{1} is based on a nonsymmetric
tensor $g_{\mu \nu }$ and affine connection $\Gamma _{\nu \rho }^{\mu }$,
defined on a non-riemannian spacetime, which may be decomposed into
symmetric and antisymmetric parts : 
\begin{eqnarray}
g_{\mu \nu } &=&g_{\left( \mu \nu \right) }+g_{\left[ \mu \nu \right] }, \\
\Gamma _{\nu \rho }^{\mu } &=&\Gamma _{\left( \nu \rho \right) }^{\mu
}+\Gamma _{\left[ \nu \rho \right] }^{\mu },
\end{eqnarray}
The contravariant tensor $g^{\mu \nu }$ is defined by : 
\begin{equation}
g^{\mu \alpha }g_{\mu \beta }=g^{\alpha \mu }g_{\beta \mu }=\delta _{\beta
}^{\alpha }.
\end{equation}
The field equations of NGT can be derived from a variational principle with
a Lagrangian density that can be defined in analogy with its counterpart in
general relativity\thinspace : 
\begin{equation}
\mathcal{L}_{NGT}=\sqrt{-g}R\equiv \sqrt{-g}g^{\mu \nu }R_{\mu \nu }\left(
W\right) ,
\end{equation}
where $g\equiv \det g_{\mu \nu }$ and $R_{\mu \nu }\left( W\right) $, the
NGT Ricci curvature tensor, is given by : 
\begin{equation}
R_{\mu \nu }\left( W\right) =\partial _{\rho }W_{\mu \nu }^{\rho }-\partial
_{\nu }W_{\mu \rho }^{\rho }+W_{\alpha \rho }^{\rho }W_{\mu \nu }^{\alpha
}-W_{\mu \rho }^{\alpha }W_{\alpha \nu }^{\rho }\,\,,
\end{equation}
in terms of the unconstrained connection coefficients $W_{\mu \nu }^{\rho }$
related to the affine connection $\Gamma _{\mu \nu }^{\rho }$ through : 
\begin{equation}
\Gamma _{\mu \nu }^{\lambda }=W_{\mu \nu }^{\lambda }+\frac{2}{3}\delta
_{\mu }^{\lambda }W_{\nu }.
\end{equation}
This last relation then implies that $\Gamma _{\mu }\equiv \Gamma _{\left[
\mu \alpha \right] }^{\alpha }=\frac{1}{2}\left( \Gamma _{\mu \alpha
}^{\alpha }-\Gamma _{\alpha \mu }^{\alpha }\right) =0$. Varying
independently the metric $g_{\mu \nu }$ and the connection $W_{\mu \nu
}^{\rho }$ in the action principle $\delta S\equiv \delta \left[ -\frac{1}{%
16\pi G}\delta \int dx^{4}\mathcal{L}_{NGT}\right] =0$, one obtains the
empty space field equations : 
\begin{eqnarray}
\partial _{\lambda }g_{\mu \nu }-\Gamma _{\mu \lambda }^{\alpha }g_{\alpha
\nu }-\Gamma _{\lambda \nu }^{\alpha }g_{\mu \alpha } &=&0, \\
\partial _{\nu }\left( \sqrt{|g|}\,g^{\left[ \mu \nu \right] }\right) &=&0,
\\
R_{\mu \nu }\left( W\right) -\frac{1}{2}g_{\mu \nu }R\left( W\right) &=&0,
\end{eqnarray}
where $R\left( W\right) \equiv g^{\mu \nu }R_{\mu \nu }\left( W\right) $. It
is worth to mention that the non trivial combination (6) is chosen so as to
yield the compatibility condition (7), which entirely determines the affine
connection $\Gamma _{\mu \nu }^{\rho }$ but not $W_{\mu \nu }^{\rho }$ \cite
{16}. In subsequent work on NGT, another choice $R_{\mu \nu }^{\prime
}\left( W\right) $ for the Ricci curvature tensor was adopted, which is : 
\begin{equation}
R_{\mu \nu }^{\prime }\left( W\right) =\partial _{\rho }W_{\mu \nu }^{\rho }-%
\frac{1}{2}\left( \partial _{\nu }W_{\mu \rho }^{\rho }+\partial _{\mu
}W_{\nu \rho }^{\rho }\right) +W_{\alpha \rho }^{\rho }W_{\mu \nu }^{\alpha
}-W_{\mu \rho }^{\alpha }W_{\alpha \nu }^{\rho }.
\end{equation}
The field equations (7)-(9) are unchanged, except that $R_{\mu \nu }\left(
W\right) $ is replaced by $R_{\mu \nu }^{\prime }\left( W\right) $. In fact
this arbitrariness is due to the absence of symmetries of the Riemann
curvature tensor of NGT : 
\begin{equation}
R^{\sigma }\,_{\mu \nu \rho }\equiv \partial _{\rho }W_{\mu \nu }^{\sigma
}-\partial _{\nu }W_{\mu \rho }^{\sigma }+W_{\alpha \rho }^{\sigma }W_{\mu
\nu }^{\alpha }-W_{\alpha \nu }^{\sigma }W_{\mu \rho }^{\alpha },
\end{equation}
which has two independant contractions $R^{\rho }\,_{\mu \nu \rho }$ and $%
R^{\rho }\,_{\rho \mu \nu }$ (the latter vanishes in general relativity).
The former contraction gives the Ricci tensor (5), while (10) corresponds to 
$R^{\rho }\,_{\mu \nu \rho }+$ $\frac{1}{8}R^{\rho }\,_{\rho \mu \nu }$ \cite
{17}. The linear approximation of masslesss NGT has been studied in ref. 
\cite{7}, where it was shown that it describes the propagation of the
massles spin 2 graviton, associated with the symmetric part of the metric,
and a massless scalar particle called skewon, associated with the
antisymmetric part of the metric. These two particles represent $2+1$
bosonic degrees of freedom.

The more recent massive version of NGT \cite{11,12} differs from the
massless one by additional non derivative terms to the Lagrangian density
(4). These are given by : 
\begin{eqnarray}
\mathcal{L}_{cosm.} &=&-2\lambda \sqrt{-g}, \\
\mathcal{L}_{skew} &=&-\frac{1}{4}\mu ^{2}\sqrt{-g}g^{\mu \nu }g_{\left[ \nu
\mu \right] }, \\
\mathcal{L}_{W} &=&-\frac{1}{6}\sqrt{-g}g^{\left( \mu \nu \right) }W_{\mu
}W_{\nu },
\end{eqnarray}
where $\lambda $ is the cosmological constant and $\mu ^{2}$ an additional
cosmological constant associated with $g_{\left[ \mu \nu \right] }$. It is
interesting to note that despite their apparent arbitrariness, these
additional terms emerge naturally in the context of non-commutative geometry 
\cite{18}. The main consequence of these additional terms is that, in the
linear approximation, the skewon becomes massive with mass $\mu $ and spin $%
1.$ Nevertheless, it is worth to mention, for future purposes, that the
geometric ( or ''kinetic'') part (4) of the Lagrangian is the same for both
massless and massive NGT, and that our arguments will equally apply in both
cases, since it is precisely based on this kinetic part.

The anholonomic formulation of NGT is based on the hypercomplexification of
the tangent space at each point $x$ of the real space-time manifold , by
allowing the functions defined on this manifold to take their values in the
ring of hyperbolic complex numbers ${H\equiv }\left\{ a+\mathrm{j}b,\left(
a,b\right) \in R^{2}\; and \;{\mathrm{j}}^{2}=+1\right\} $ ($\mathrm{j}%
^{2}=-1$ in the case of the field of ordinary complex numbers) \cite{4}.
This leads to the introduction of an hyperbolic complex vierbein $e_{\mu
}^{a}$ ($\mu =\overline{1,4}$ ; $a=\overline{1,4}$) \cite{6} : 
\begin{equation}
e_{\mu }^{a}=a_{\mu }^{a}+\mathrm{j}b_{\mu }^{a},
\end{equation}
where $a_{\mu }^{a}$ and $b_{\mu }^{a}$ are real valued. The hyperbolic
complex conjugate (hereafter abbreviated as H.C.C.) vierbein is $\widetilde{e%
}_{\mu }^{a}=a_{\mu }^{a}-\mathrm{j}b_{\mu }^{a}$. One also introduces the
inverse vierbein $e_{a}^{\mu }$, orthonormal to $e_{\mu }^{a}$ and thus : $%
e_{\mu }^{a}e_{b}^{\mu }=\delta _{b}^{a}$ and $e_{\mu }^{a}e_{a}^{\nu
}=\delta _{\mu }^{\nu }$ (idem. for H.C.C.).

The hyperbolic complex sesquilinear form $^{c}g_{\mu \nu }$ (and its inverse 
$^{c}g^{\mu \nu }$) acting on the tangent space is given by : 
\begin{eqnarray}
^{c}g_{\mu \nu } &=&e_{\mu }^{a}\widetilde{e}_{\nu }^{b}\eta _{ab}, \\
^{c}g^{\mu \nu } &=&e_{a}^{\mu }\widetilde{e}_{b}^{\nu }\eta ^{ab},
\end{eqnarray}
where $\eta _{ab}=diag\left( +1,-1,-1,-1\right) $ is the Minkowski flat
space-time metric. As $\eta _{ab}$ is symmetric, we have $^{c}\widetilde{g}%
_{\mu \nu }=\left. ^{c}g_{\nu \mu }\right. $, where $^{c}\widetilde{g}_{\mu
\nu }$ is the H.C.C. of $^{c}g_{\mu \nu }$. This means that $^{c}g_{\mu \nu
} $ can be decomposed as follows : 
\begin{equation}
^{c}g_{\mu \nu }=g_{\left( \mu \nu \right) }+\mathrm{j}g_{\left[ \mu \nu %
\right] },
\end{equation}
where $g_{\left( \mu \nu \right) }$ and $g_{\left[ \mu \nu \right] }$ are
respectively the symmetric and antisymmetric parts of the real metric
tensor. The group of local isometries that preserves the metric is the group
of hyperbolic complex valued transformations $U^{a}\,_{b}\left( x\right)
e_{\mu }^{b}\left( x\right) $ acting on the tangent space such that $%
^{c}g_{\mu \nu }^{\prime }\left( x\right) =\left. ^{c}g_{\mu \nu }\left(
x\right) \right. $, where : 
\begin{equation}
^{c}g_{\mu \nu }^{\prime }\left( x\right) \equiv e_{\mu }^{\prime a}\left(
x\right) \widetilde{e}_{\nu }^{\prime b}\left( x\right) \eta
_{ab}=U^{a}\,_{c}\left( x\right) \widetilde{U}^{b}\,_{d}\left( x\right) \eta
_{ab}e_{\mu }^{c}\left( x\right) \widetilde{e}_{\nu }^{d}\left( x\right) \,.
\end{equation}
Comparing (16) with (19), one then deduces that the transformation $U$ must
satisfy the relation : 
\begin{equation}
\eta =U^{+}\eta U,
\end{equation}
where $U^{+}\equiv \widetilde{U}^{T}$. This is the defining relation for the
pseudo-unitary group (in the sens of hyperbolic complex numbers) $U(3,1,{H)}$%
, which has been shown to be isomorphic to $GL\left( 4,R\right) $ \cite{5}.
It is precisely this isomorphism that selects the hyperbolic complex
numbers. Had one chosen ordinary complex numbers, then one would have ended
with $U(3,1,{C)}$ wich is not isomorphic to $GL\left( 4,R\right) $. The
group $U(3,1,{H)}$ plays the same role in NGT as does the local homogeneous
Lorentz group $SO\left( 3,1\right) $ in general relativity.

To each of the sixteen generators of the local group of invariance $U(3,1,{H)%
}$ corresponds a compensating field, namely the hyperbolic complex spin
connection $^{c}\omega _{\mu }\,^{ab}$, with the inhomogeneous
transformation law : 
\begin{equation}
^{c}\omega _{\mu }^{\prime }\,^{a}\,_{b}=U^{a}\,_{d}\left( ^{c}\omega _{\mu
}\,^{d}\,_{f}\right) \left( U^{-1}\right) ^{f}\,_{b}-\left( \partial _{\mu
}U\right) ^{a}\,_{d}U^{d}\,_{b}.
\end{equation}
In accordance with (20), $^{c}\omega _{\mu }\,^{ab}$ satisfies the relation $%
^{c}\widetilde{\omega }_{\mu }\,^{ab}=-\left. ^{c}\omega _{\mu
}\,^{ba}\right. $. This property then implies the decomposition into
antisymmetric ``real'' part and symmetric ``imaginary'' part : 
\begin{equation}
^{c}\omega _{\mu }\,^{ab}=\omega _{\mu }\,^{\left[ ab\right] }+\mathrm{j}%
\omega _{\mu }\,^{\left( ab\right) }.
\end{equation}
This spin connection enters the definition of the covariant derivative,
which acts on hyperbolic complex vectors $V^{a}$ defined on the tangent
space : 
\begin{equation}
^{c}\mathcal{D}_{\mu }V^{a}=\partial _{\mu }V^{a}+\left. ^{c}\omega _{\mu
}\,^{a}\,_{b}V^{b}\right. .
\end{equation}
The anholonomic curvature tensor can then be introduced through the
commutator of two covariant derivatives : 
\begin{equation}
\left[ ^{c}\mathcal{D}_{\mu },^{c}\mathcal{D}_{\nu }\right] ^{a}\,_{b}\equiv
R_{\mu \nu }\,^{a}\,_{b},
\end{equation}
where : 
\begin{equation}
R_{\mu \nu }\,^{ab}=\partial _{\mu }\left. ^{c}\omega _{\nu }\,^{ab}\right.
-\partial _{\nu }\left. ^{c}\omega _{\mu }\,^{ab}\right. +\left. ^{c}\omega
_{\mu }\,^{a}\,_{d}\right. \left. ^{c}\omega _{\nu }\,^{db}\right. -\left.
^{c}\omega _{\nu }\,^{a}\,_{d}\right. \left. ^{c}\omega _{\mu
}\,^{db}\right. .
\end{equation}
To recover the Lagrangian (4) in terms of anholonomic quantities, one has to
impose a compatibility condition between the connections $W_{\mu \nu }^{\rho
}$ and $^{c}\omega _{\mu }\,^{ab}$. This condition is given by \cite{6,17} : 
\begin{equation}
\partial _{\sigma }e_{\mu }^{a}+\left. ^{c}\omega _{\sigma
}\,^{a}\,_{b}\right. e_{\mu }^{b}-\left. ^{c}W_{\mu \sigma }^{\rho }\right.
e_{\rho }^{a}=0\,\,,
\end{equation}
where we have introduced the hyperbolic complex valued connection : 
\begin{equation}
\left. ^{c}W_{\mu \sigma }^{\rho }\right. \equiv Re\left( ^{c}W_{\mu \sigma
}^{\rho }\right) +\mathrm{j}Im\left( ^{c}W_{\mu \sigma }^{\rho }\right) .
\end{equation}
After solving (26) for $^{c}\omega _{\sigma }\,^{a}\,_{b}\left( e_{\mu
}^{a},^{c}W_{\mu \sigma }^{\rho }\right) $ and substituting in (25), one can
relate holonomic to anholonomic tensors : 
\begin{eqnarray}
R_{\mu \nu }\,^{a}\,_{b}e_{a}^{\lambda }e_{\sigma }^{b} &=&\left.
^{c}R^{\lambda }\,_{\sigma \nu \mu }\left( ^{c}W_{\beta \rho }^{\alpha
}\right) \right. \,\,, \\
e_{\mu }^{b}\left( e_{a}^{\alpha }R_{\alpha \nu }\,^{a}\,_{b}\right)
&=&\left. ^{c}R_{\mu \nu }\left( ^{c}W_{\beta \rho }^{\alpha }\right)
\right. \,\,, \\
e^{\mu }\,_{a}R_{\mu }\,^{a} &\equiv &e_{a}^{\mu }\left( \widetilde{e}%
_{b}^{\nu }R_{\mu \nu }\,^{ab}\right) =\left. ^{c}g^{\mu \nu }\right. \left.
^{c}R_{\mu \nu }\left( ^{c}W_{\beta \rho }^{\alpha }\right) \right. =R\,\,.\,
\end{eqnarray}
The hyperbolic complex Riemann and Ricci tensors, $^{c}R^{\lambda
}\,_{\sigma \nu \mu }\left( ^{c}W_{\beta \rho }^{\alpha }\right) $ and $%
^{c}R_{\mu \nu }\left( ^{c}W_{\beta \rho }^{\alpha }\right) $, have the same
structure as their real counterparts (11) and (5), with the connection $%
W_{\mu \sigma }^{\rho }$ replaced by the hyperbolic complex one defined in
(27). Most important is the last relation (30) : it defines the Ricci scalar
which is \emph{real} and coincides with the one that enters the definition
(4) of $\mathcal{L}_{NGT}$ \cite{17}. It follows that the real NGT
Lagrangian density (4) can be written in terms of anholonomic quantities as
: 
\begin{equation}
\mathcal{L}_{NGT}=\left( e\widetilde{e}\right) ^{\frac{1}{2}}e_{a}^{\mu }%
\widetilde{e}_{b}^{\nu }R_{\mu \nu }\,^{ab}\left( ^{c}\omega \right) ,
\end{equation}
where $e\equiv \det e_{\mu }^{a}$ , $\widetilde{e}\equiv \det \widetilde{e}%
_{\mu }^{a}$ and $\left( e\widetilde{e}\right) =\left| ^{c}g_{\mu \nu
}\right| \in R$. This is the Lagrangian that we will need for our model of
the following section.

We note, for completeness, that the Lagrangian density (4) with the choice
(10) for $R_{\mu \nu }\left( W\right) $ can also be written in the anholomic
formalism as \cite{17} : 
\begin{equation}
\mathcal{L}_{NGT}=\left( e\widetilde{e}\right) ^{\frac{1}{2}}\left[
e_{a}^{\mu }\widetilde{e}_{b}^{\nu }R_{\mu \nu }\,^{ab}\left( ^{c}\omega
_{\alpha }\right) +\frac{1}{8}e_{b}^{\mu }\widetilde{e}^{\nu b}R_{\mu \nu
}\,^{a}\,_{a}\left( ^{c}\omega _{\alpha }\right) \right] .
\end{equation}

\section{The model}

Let us first state more precisely our goal. We want to investigate the
possibility of constructing a minimal model of a locally supersymmetric
extension of NGT, that is a field theory containing NGT and invariant under
local supersymmetric transformations, a kind of nonsymmetric simple
supergravity where NGT would replace general relativity. We want furthermore
to work in the framework of the superPoincar\'e group, the representations
of which are well known. In this connection, we must associate to each
bosonic particle its fermionic partner. In the context of massless NGT (to
which we restrict our investigation), one must have two fermionic fields : a
spin $\frac{3}{2}$ gravitino $\psi _{\mu }\equiv e_{\mu }^{a}\psi _{a}$,
where $e_{\mu }^{a}$ is the hyperbolic complex vierbein and $\psi _{a}$ a
non-hyperbolic complex vectorial spinor, and a (non-hyperbolic complex) spin 
$\frac{1}{2}$ skewino $\chi $. As we work with Majorana spinors, these two
fields represent four fermionic degrees of freedom (in $d=4$ dimensions).
However, the graviton and skewon represent only three bosonic degrees of
freedom. There is thus a mismatch of bosonic and fermionic degrees of
freedom. To overcome this difficulty one has to introduce a second scalar
field $\varphi $, thereby completing two representations of the
superPoincar\'e group : one contains the spin $2$ graviton and the spin $%
\frac{3}{2}$ gravitino, the other contains the skewon and the additional
field $\varphi $ (two scalar fields) and the spin $\frac{1}{2}$ skewino.

By minimal, we mean that NGT enters the model through the Lagrangian (31),
without additional terms containing the derivative of the spin connection
(22). Insofar as we are concerned with representations of the
superPoincar\'{e} group, we introduce the $SO(3,1)$ covariant derivative $%
D_{\nu }$ acting on the spinors $\psi _{\mu }$ and $\chi $ : 
\begin{equation}
D_{\nu }\equiv \partial _{\nu }+\frac{1}{2}\omega _{\nu }\,^{\left[ ab\right]
}\sigma _{ab},
\end{equation}
where $\omega _{\nu }\,^{\left[ ab\right] }$ is the antisymmetric real part
of the hyperbolic complex spin connection (22), and $\sigma _{ab}\equiv 
\frac{1}{4}\left[ \gamma _{a},\gamma _{b}\right] $, $\gamma _{a}$ being the
constant Dirac matrices. This derivative is covariant under the restriction
of the local group $U(3,1,{H)}$ to its real unimodular subgroup, which is
precisely the homogeneous Lorentz group $SO(3,1)$. To accomodate this
privileged role of $\omega _{\mu }\,^{\left[ ab\right] }$, we split the
Ricci tensor $R_{\mu \nu }\,^{ab}$ and the Lagrangian $\mathcal{L}_{NGT}$
(31) into two parts, one of which contains only the Lorentz antisymmetric
part $\omega _{\mu }\,^{\left[ ab\right] }$. We thus get $R_{\mu \nu
}\,^{ab}=R_{\mu \nu }^{1}\,^{ab}+R_{\mu \nu }^{2}\,^{ab}$ where : 
\begin{eqnarray}
R_{\mu \nu }^{1}\,^{ab} &=&\partial _{\mu }\omega _{\mu }\,^{\left[ ab\right]
}+\omega _{\mu }\,^{[a}\,_{c]}\omega _{\nu }^{\left[ cb\right] }-\left( \mu
\longleftrightarrow \nu \right) \,, \\
R_{\mu \nu }^{2}\,^{ab} &=&\mathrm{j}D_{\mu }\omega _{\nu }\,^{\left(
ab\right) }+\omega _{\mu }\,^{(a}\,_{c)}\omega _{\nu }^{\left( cb\right)
}-\left( \mu \longleftrightarrow \nu \right) \,,
\end{eqnarray}
with $D_{\mu }\omega _{\nu }\,^{\left( ab\right) }=\partial _{\mu }\omega
_{\nu }\,^{\left( ab\right) }+\omega _{\mu }\,^{[a}\,_{c]}\omega _{\nu
}\,^{\left( cb\right) }+\omega _{\mu }\,^{[b}\,_{c]}\omega _{\nu }\,^{\left(
ac\right) }\,$\footnote{%
This relation is to be inderstood as a mere condensed notation ( $D_{\mu
}\omega _{\nu }\,^{\left( ab\right) }$ is not covariant under $SO(3,1),$
because $\omega _{\nu }\,^{\left( ab\right) }$ transforms as a connection).}$%
.$ By inserting this decomposition into (31), we get : 
\begin{equation}
\mathcal{L}_{NGT}=\mathcal{L}^{1}+\mathcal{L}^{2},
\end{equation}
where : 
\begin{eqnarray}
\mathcal{L}^{1} &=&\left( e\widetilde{e}\right) ^{\frac{1}{2}}e_{a}^{\mu }%
\widetilde{e}_{b}^{\nu }R_{\mu \nu }^{1}\,^{ab}\,, \\
\mathcal{L}^{2} &=&\left( e\widetilde{e}\right) ^{\frac{1}{2}}e_{a}^{\mu }%
\widetilde{e}_{b}^{\nu }R_{\mu \nu }^{2}\,^{ab}\,.
\end{eqnarray}
The relevance of this decomposition will appear shortly.

For the gravitino field $\psi _{\mu }$, we choose to generalize the minimal
coupling of the Rarita-Schwinger field to ordinary gravity \cite{13}, by
merely replacing the real vierbein of general relativity by its NGT
hyperbolic complex counterpart. We thus put : 
\begin{equation}
\mathcal{L}_{\psi }=k\Omega ^{\mu \nu \alpha \rho }\overline{\psi }_{\mu
}\gamma _{5}\gamma _{\nu }D_{\alpha }\psi _{\rho }+H.C.C.\,,
\end{equation}
where $k$ is a numerical factor to be fixed later. $\Omega ^{\mu \nu \alpha
\rho }$ is some linear combination of terms of the form $\left( e\widetilde{e%
}\right) ^{\frac{1}{2}}\eta ^{abcd}a_{a}^{\mu }a_{b}^{\nu }a_{c}^{\alpha
}a_{d}^{\rho }$, where $\eta ^{abcd}$ is the completely antisymmetric symbol
and each factor $a_{i}^{\sigma }$ ($\sigma =\mu ,\nu ,\alpha ,\rho $ ; $%
i=a,b,c,d$ ) is either $e_{i}^{\sigma }$ or $\widetilde{e}_{i}^{\sigma }$.
The matrices $\gamma _{\nu }$ and $\gamma _{5}$ are defined by $\gamma _{\nu
}\equiv e_{\nu }^{a}\gamma _{a}$ ($\gamma _{a}$ constant Dirac matrix) and $%
\gamma _{5}\equiv \imath \gamma _{0}\gamma _{1}\gamma _{2}\gamma _{3}$ ($%
\imath ^{2}=-1$). The covariant derivative $D_{\alpha }\psi _{\rho }$ is
given by : 
\begin{equation}
D_{\alpha }\psi _{\rho }=\partial _{\alpha }\psi _{\rho }+\frac{1}{2}\omega
_{\alpha }\,^{\left[ ab\right] }\sigma _{ab}\psi _{\rho },
\end{equation}
and $\overline{\psi }_{\mu }\equiv e_{\mu }^{a}\psi _{a}^{T}C$, $C$ being
the charge conjugation matrix ($C\gamma _{a}C^{-1}=-\gamma _{a}^{T}$, $%
\,C^{T}=-C$). It must be stressed that the use of the superPoincar\'{e}
representations entails the introduction of this covariant derivative, due
to the antisymmetry of the $SO(3,1)$ generators $\sigma _{ab}$. Furthermore,
the commutator of two derivatives generates the tensor $R_{\mu \nu
}^{1}\,^{ab}$ (34) : 
\begin{equation}
\left[ D_{\mu },D_{\nu }\right] \psi _{\rho }=\frac{1}{2}R_{\mu \nu
}^{1}\,^{ab}\sigma _{ab}\psi _{\rho },
\end{equation}
hence the decomposition (37)-(38).

We likewise take the minimally coupled Lagrangians for the skewino $\chi $
and the scalar field $\varphi $, namely : 
\begin{eqnarray}
\mathcal{L}_{\chi } &=&-\frac{1}{2}\overline{\chi }\,D\mathcal{\!}%
\!\!\!/\chi \,, \\
\mathcal{L}_{\varphi } &=&\frac{1}{2}g^{\mu \nu }\partial _{\mu }\varphi
\partial _{\nu }\varphi =\frac{1}{2}g^{\left( \mu \nu \right) }\partial
_{\mu }\varphi \partial _{\nu }\varphi \,,
\end{eqnarray}
where : 
\begin{eqnarray}
D\mathcal{\!}\!\!\!/\chi &\equiv &\gamma ^{\mu }D_{\mu }\chi =\gamma ^{\mu
}\left( \partial _{\mu }\chi +\frac{1}{2}\omega _{\mu }\,^{\left[ ab\right]
}\sigma _{ab}\chi \right) \,, \\
g^{\left( \mu \nu \right) } &\equiv &\frac{1}{2}\left( e_{a}^{\mu }%
\widetilde{e}_{b}^{\nu }+e_{a}^{\nu }\widetilde{e}_{b}^{\mu }\right) \eta
^{ab}\,.
\end{eqnarray}
Although not necessary for our line of argument, we give them for
completeness.

Having postulated the Lagrangians of the fields that enter our model, we
will show in the following section that, with the simplest choice of
supersymmetric variation of the vierbein $e_{\mu }^{a}$ and gravitino $\psi
_{\mu }$, it is impossible to fix the tensor $\Omega ^{\mu \nu \alpha \rho }$
so that the linear contributions in $\psi _{\mu }$ coming from variations of 
$\mathcal{L}_{NGT}$ and $\mathcal{L}_{\psi }$ cancel.

We finally note that, in order to simplify matters, we will use the 1.5
order formalism \cite{19}. The spin connection $^{c}\omega _{\mu }\,^{ab}$
is considered as an independant field obeying an algebraic equation which
gives its expression in terms of the vierbein and the spinorial fields. This
equation is obtained by varying the spin connection $^{c}\omega _{\mu
}\,^{ab}$ in the action, as an independent field (or the real and imaginary
parts $\omega _{\mu }\,^{\left[ ab\right] }$ and $\omega _{\mu }\,^{\left(
ab\right) }$ as independent fields) and is given by : 
\begin{eqnarray}
D_{\mu }\left[ \left( e\widetilde{e}\right) ^{\frac{1}{2}}\left( e_{a}^{\mu }%
\widetilde{e}_{b}^{\nu }-\widetilde{e}_{b}^{\mu }e_{a}^{\nu }\right) \right]
&=&\mathrm{j}\left( e\widetilde{e}\right) ^{\frac{1}{2}}\left[ \omega _{\mu
}\,_{(a}\,^{c)}\left( e_{c}^{\mu }\widetilde{e}_{b}^{\nu }-\widetilde{e}%
_{b}^{\mu }e_{c}^{\nu }\right) \right.  \nonumber \\
&&\left. +\omega _{\mu }\,_{(b}\,^{c)}\left( \widetilde{e}_{c}^{\mu
}e_{a}^{\nu }-e_{a}^{\mu }\widetilde{e}_{c}^{\nu }\right) \right] -S_{\left[
ab\right] }^{\nu }\,,
\end{eqnarray}
where $S_{\left[ ab\right] }^{\nu }$ is defined by $\delta \left( \mathcal{L}%
_{\psi }+\mathcal{L}_{\chi }\right) =S_{\left[ ab\right] }^{\nu }\delta
\omega _{\mu }\,^{\left[ ab\right] }$. Most important in the 1.5 order
formalism is that the spin connection is not varied when we come to the
local supersymmetric variation of the Lagrangians.

\section{Terms linear in the gravitino field}

Our strategy is to first investigate the piece of the supersymmetric
variation of $\mathcal{L}_{NGT}$ and $\mathcal{L}_{\psi }$ which is linear
in the gravitino field $\psi _{\mu }$. To clearly state our line of
argument, it is convenient to first write down the variation of $\mathcal{L}%
^{1}$ (37), under an infinitesimal variation of the hyperbolic complex
vierbein $\delta e_{\mu }^{a}$. This variation can be put into the form : 
\begin{equation}
\delta \mathcal{L}^{1}=\frac{1}{2}\left( e\widetilde{e}\right) ^{\frac{1}{2}}%
\left[ e_{i}^{\rho }R^{1}-2e_{i}^{\mu }e_{a}^{\rho }R_{\mu }^{1}\,^{a}\right]
\delta e_{\rho }^{i}+H.C.C.,
\end{equation}
where $R_{\mu }^{1}\,^{a}\equiv \widetilde{e}_{b}^{\nu }R_{\mu \nu
}^{1}\,^{ab}$ and $R^{1}\equiv e_{a}^{\mu }R_{\mu }^{1}\,^{a}$. We have used
the relation $\delta \left( e\widetilde{e}\right) ^{\frac{1}{2}}=\frac{1}{2}%
\left( e\widetilde{e}\right) ^{\frac{1}{2}}e_{a}^{\mu }\delta e_{\mu
}^{a}+H.C.C.$ and the symmetries of $R_{\mu \nu }^{1}\,^{ab}$ ($R_{\mu \nu
}^{1}\,^{ab}=-R_{\mu \nu }^{1}\,^{ba}=R_{\nu \mu }^{1}\,^{ba}$).
Generalizing the supersymmetric variation of the tetrad of general
relativity in the context of ordinary supergravity, to the hyperbolic
complex vierbein of NGT, we put : 
\begin{eqnarray}
\delta _{0}e_{\mu }^{a} &=&\overline{\varepsilon }\gamma ^{a}\psi _{\mu
}\equiv \overline{\varepsilon }\gamma ^{a}\psi _{b}e_{\mu }^{b}, \\
\delta _{0}\widetilde{e}_{\mu }^{a} &=&\overline{\varepsilon }\gamma ^{a}%
\widetilde{\psi }_{\mu }\equiv \overline{\varepsilon }\gamma ^{a}\psi _{b}%
\widetilde{e}_{\mu }^{b},
\end{eqnarray}
where $\varepsilon \left( x\right) $ is an infinitesimal spin $\frac{1}{2}$
Majorana spinor, the parameter of the local supersymmetric transformation.
Upon substitution in (47), one gets the term linear in $\psi _{\mu }$ : 
\begin{eqnarray}
\delta \mathcal{L}_{0}^{1} &=&\frac{1}{2}\left( e\widetilde{e}\right) ^{%
\frac{1}{2}}\left[ e_{i}^{\rho }R^{1}-2e_{i}^{\mu }e_{a}^{\rho }R_{\mu
}^{1}\,^{a}\right] \left( \overline{\varepsilon }\gamma ^{i}\psi _{j}\right)
e_{\rho }^{j}+H.C.C.  \nonumber \\
&=&\frac{1}{2}\left( e\widetilde{e}\right) ^{\frac{1}{2}}\left[ R^{1}\left( 
\overline{\varepsilon }\gamma ^{i}\psi _{i}\right) -2R_{\mu
}^{1}\,^{a}\left( \overline{\varepsilon }\gamma ^{\mu }\psi _{a}\right) %
\right] +H.C.C.\,.
\end{eqnarray}
The particularity of this term is first that it is linear in $\psi _{\mu }$,
and second that it contains contractions of the tensor $R_{\mu \nu
}^{1}\,^{ab}$ with $e_{\mu }^{a}$ and/or $\widetilde{e}_{\mu }^{a}$. It is a
common feature, if not a rule, of supergravity theories that this term is
cancelled by the variation of the gravitino Lagrangian generated by the part
of the infinitesimal variation : 
\begin{equation}
\delta _{0}\psi _{\mu }\sim D_{\mu }\varepsilon =\partial _{\mu }\varepsilon
+\frac{1}{2}\omega _{\mu }\,^{ab}\sigma _{ab}\varepsilon .
\end{equation}
This situation is due to the fact that the Rarita-Schwinger-like Lagrangian
is the only term that can yield, upon the variation $\delta _{0}\psi _{\mu }$
a term linear in $\psi _{\mu }$ and containing the Ricci tensor, which
appears via the commutator of two covariant derivatives. Similarly, in our
case, the only term that might cancel the contribution (50) is the
contribution linear in $\psi _{\mu }$ coming from the variation of $\mathcal{%
L}_{\psi }$ (see (39)), with $\delta _{0}\psi _{\mu }=D_{\mu }\varepsilon $
(a possible numerical factor can be absorbed in the constant $k$ appearing
in (39)). Thus, the problem amounts, at the first stage, to find, if it
exists, the form of the tensor $\Omega ^{\mu \nu \alpha \rho }$, so as to
cancel the contribution (50).

Taking $\delta _{0}\psi _{\mu }=D_{\mu }\varepsilon $, the variation $\delta 
\mathcal{L}_{\psi }^{0}$ of $\mathcal{L}_{\psi }$ that follows is : 
\begin{eqnarray}
\delta \mathcal{L}_{\psi }^{0} &=&\left\{ k\Omega ^{\mu \nu \alpha \rho
}\left( \delta _{0}\overline{\psi }_{\mu }\right) \gamma _{5}\gamma _{\nu }%
\mathcal{D}_{\alpha }\psi _{\rho }\right.  \nonumber \\
&&\quad \quad \quad \quad \quad \quad \left. +k\Omega ^{\mu \nu \alpha \rho }%
\overline{\psi }_{\mu }\gamma _{5}\gamma _{\nu }\mathcal{D}_{\alpha }\left(
\delta _{0}\psi _{\rho }\right) \right\} +H.C.C.  \nonumber \\
&=&\left\{ k\Omega ^{\mu \nu \alpha \rho }\left( \mathcal{D}_{\mu }\overline{%
\varepsilon }\right) \gamma _{5}\gamma _{\nu }\mathcal{D}_{\alpha }\psi
_{\rho }\right.  \nonumber \\
&&\quad \quad \quad \quad \quad \quad \left. +k\Omega ^{\mu \nu \alpha \rho }%
\overline{\psi }_{\mu }\gamma _{5}\gamma _{\nu }\mathcal{D}_{\alpha }\left( 
\mathcal{D}_{\rho }\varepsilon \right) \right\} +H.C.C.\,.
\end{eqnarray}
After partially integrating the first term, and omitting the terms
containing $D_{\mu }\left( \Omega ^{\mu \nu \alpha \rho }\right) $ and $%
D_{\mu }\gamma _{\nu }$, we obtain two terms that contain a double covariant
derivative (from which one can generate $R_{\mu \nu }^{1}\,^{ab}$) and are
linear in $\psi _{\mu }$ : 
\begin{equation}
\delta \mathcal{L}_{\psi }^{0^{\prime }}=\left\{ -k\Omega ^{\mu \nu \alpha
\rho }\overline{\varepsilon }\gamma _{5}\gamma _{\nu }\mathcal{D}_{\mu }%
\mathcal{D}_{\alpha }\psi _{\rho }+k\Omega ^{\mu \nu \alpha \rho }\overline{%
\psi }_{\mu }\gamma _{5}\gamma _{\nu }\mathcal{D}_{\alpha }\mathcal{D}_{\rho
}\varepsilon \right\} +H.C.C.\,.
\end{equation}
These two terms are the only ones that might cancel the contribution (50),
whatever the complete laws of transformation of the fields may be. Indeed,
the terms that one can add to the transformation law (51) are at least
linear in $\psi _{\mu }$, yielding quadratic terms in the variation of $%
\mathcal{L}_{\psi }$, and the linear part of the transformation laws of the
fields $\chi $ and $\varphi $ : 
\begin{eqnarray}
\delta \chi &=&a_{1}\gamma ^{\mu }\mathcal{D}_{\mu }\left( \varphi
\varepsilon \right) +a_{2}\left( \mathcal{D}_{\mu }\gamma ^{\mu }\right)
\varphi \varepsilon , \\
\delta \varphi &=&b_{1}\overline{\varepsilon }\chi +b_{2}\overline{%
\varepsilon }\gamma ^{\mu }\psi _{\mu },
\end{eqnarray}
($a_{1},a_{2},b_{1}$ and $b_{2}$ are constants) cannot contribute to the
cancellation of (50).

On the other hand, to cancel (50) $\delta \mathcal{L}_{\psi }^{0^{\prime }}$
must be of the form : 
\begin{eqnarray}
\delta \mathcal{L}_{\psi }^{0^{\prime }} &=&\left\{ -k\left( e\widetilde{e}%
\right) ^{\frac{1}{2}}\left( \eta ^{\nu \rho \alpha \widetilde{\mu }}+\eta
^{\nu \rho \widetilde{\alpha }\mu }\right) \overline{\varepsilon }\gamma
_{5}\gamma _{\nu }\mathcal{D}_{\mu }\mathcal{D}_{\alpha }\psi _{\rho }\right.
\nonumber \\
&&\left. +k\left( e\widetilde{e}\right) ^{\frac{1}{2}}\left( \eta ^{\mu \nu
\alpha \widetilde{\rho }}+\eta ^{\mu \nu \widetilde{\alpha }\rho }\right) 
\overline{\psi }_{\mu }\gamma _{5}\gamma _{\nu }\mathcal{D}_{\alpha }%
\mathcal{D}_{\rho }\varepsilon \right\} +H.C.C.\,,
\end{eqnarray}
where 
\begin{eqnarray}
\eta ^{\nu \rho \alpha \widetilde{\mu }} &=&\eta ^{abcd}e_{a}^{\nu
}e_{b}^{\rho }e_{c}^{\alpha }\widetilde{e}_{d}^{\mu }\,, \\
\eta ^{\nu \rho \widetilde{\alpha }\mu } &=&\eta ^{abcd}e_{a}^{\nu
}e_{b}^{\rho }\widetilde{e}_{c}^{\alpha }e_{d}^{\mu }\,, \\
\eta ^{\mu \nu \alpha \widetilde{\rho }} &=&\eta ^{abcd}e_{a}^{\mu
}e_{b}^{\nu }e_{c}^{\alpha }\widetilde{e}_{d}^{\rho }\,, \\
\eta ^{\mu \nu \widetilde{\alpha }\rho } &=&\eta ^{abcd}e_{a}^{\mu
}e_{b}^{\nu }\widetilde{e}_{c}^{\alpha }e_{d}^{\rho }
\end{eqnarray}
($\eta ^{abcd}$ is the completely antisymmetric symbol). The form (56) of $%
\delta \mathcal{L}_{\psi }^{0^{\prime }}$ is dictated by two requirements :
i) the $\eta $ factors must have the right (anti)symmetry properties as to
yield the commutator of two covariant derivatives, and hence $R_{\mu \nu
}^{1}\,^{ab}$, and ii) once the Ricci tensor $R_{\mu \nu }^{1}\,^{ab}$
generated, it must have $e_{\bullet }^{\mu }\widetilde{e}_{\bullet }^{\nu }$
( or $\widetilde{e}_{\bullet }^{\mu }e_{\bullet }^{\nu }$ ) in front of it,
in order to get the same contractions as in (50) and cancel this term.

Let us briefly outline how (56) cancels (50). By relabeling indices of the $%
\eta $ terms in (56), and taking into account the expression (41), one gets
: 
\begin{eqnarray}
\delta \mathcal{L}_{\psi }^{0^{\prime }} &=&-k\left( e\widetilde{e}\right) ^{%
\frac{1}{2}}\left\{ \eta ^{\nu \rho \alpha \widetilde{\mu }}\overline{%
\varepsilon }\gamma _{5}\gamma _{\nu }\left[ \mathcal{D}_{\mu },\mathcal{D}%
_{\alpha }\right] \psi _{\rho }\right.  \nonumber \\
&&\,\,\quad \quad \quad \quad \quad \quad \quad \left. -\eta ^{\mu \nu
\alpha \widetilde{\rho }}\overline{\psi }_{\mu }\gamma _{5}\gamma _{\nu }%
\left[ \mathcal{D}_{\alpha },\mathcal{D}_{\rho }\right] \varepsilon \right\}
+H.C.C.  \nonumber \\
&=&-\frac{k}{2}\left( e\widetilde{e}\right) ^{\frac{1}{2}}\eta ^{\nu \rho
\alpha \widetilde{\mu }}\left[ \left( \overline{\varepsilon }\gamma
_{5}\gamma _{\nu }\sigma _{ab}\psi _{\rho }\right) \right.  \nonumber \\
&&\quad \quad \quad \quad \quad \quad \quad \quad \left. -\left( \overline{%
\psi }_{\rho }\gamma _{5}\gamma _{\nu }\sigma _{ab}\varepsilon \right) %
\right] R_{\mu \alpha }^{1}\,^{ab}\,+H.C.C..
\end{eqnarray}
Using the properties of the charge conjugation matrix $C$ and Dirac matrices
(especially that of the anticommutator $\left\{ \sigma _{ab},\gamma
_{c}\right\} =\imath \eta _{abcd}\gamma _{5}\gamma ^{d}$), the last
expression transforms to : 
\begin{eqnarray}
\delta \mathcal{L}_{\psi }^{0^{\prime }} &=&-\imath \frac{k}{2}\left( e%
\widetilde{e}\right) ^{\frac{1}{2}}\eta ^{cijk}\eta _{cabd}e_{i}^{\rho
}e_{j}^{\alpha }\widetilde{e}_{k}^{\mu }\left( \overline{\varepsilon }\gamma
^{d}\psi _{\rho }\right) R_{\mu \alpha }^{1}\,^{ab}+H.C.C.  \nonumber \\
&=&\imath \frac{k}{2}\left( e\widetilde{e}\right) ^{\frac{1}{2}}\delta _{%
\left[ abd\right] }^{\left[ ijk\right] }\,e_{i}^{\rho }e_{j}^{\alpha }%
\widetilde{e}_{k}^{\mu }\left( \overline{\varepsilon }\gamma ^{d}\psi _{\rho
}\right) R_{\mu \alpha }^{1}\,^{ab}\,+H.C.C.,
\end{eqnarray}
with 
\begin{equation}
\delta _{\left[ abd\right] }^{\left[ ijk\right] }\equiv \delta
_{a}^{i}\delta _{b}^{j}\delta _{d}^{k}-\delta _{a}^{i}\delta _{d}^{j}\delta
_{b}^{k}+\delta _{b}^{i}\delta _{d}^{j}\delta _{a}^{k}-\delta _{b}^{i}\delta
_{a}^{j}\delta _{d}^{k}+\delta _{d}^{i}\delta _{a}^{j}\delta _{b}^{k}-\delta
_{d}^{i}\delta _{b}^{j}\delta _{a}^{k}\,.
\end{equation}
Using the symmetry properties of $R_{\mu \nu }^{1}\,^{ab}$ ($R_{\mu \nu
}^{1}\,^{ab}=-R_{\nu \mu }^{1}\,^{ab}=-R_{\mu \nu }^{1}\,^{ba}=R_{\nu \mu
}^{1}\,^{ba}$) along with the definitions $R_{\mu }^{1}\,^{a}=\widetilde{e}%
_{b}^{\nu }R_{\mu \nu }^{1}\,^{ab}$, $\widetilde{R_{\nu }^{1}\,^{b}}%
=e_{a}^{\mu }R_{\mu \nu }^{1}\,^{ab}$ and $R^{1}=e_{a}^{\mu }\widetilde{e}%
_{b}^{\nu }R_{\mu \nu }^{1}\,^{ab}=\widetilde{R^{1}}$ , we end up with the
following expression : 
\begin{eqnarray}
\delta \mathcal{L}_{\psi }^{0^{\prime }} &=&\left\{ -\imath \frac{k}{2}%
\left( e\widetilde{e}\right) ^{\frac{1}{2}}\left[ R^{1}\left( \overline{%
\varepsilon }\gamma ^{d}\psi _{d}\right) -2R_{\mu }^{1}\,^{a}\left( 
\overline{\varepsilon }\gamma ^{\mu }\psi _{a}\right) \right] \right. 
\nonumber \\
&&\left. -\imath \frac{k}{2}\left( e\widetilde{e}\right) ^{\frac{1}{2}}\left[
R^{1}\left( \overline{\varepsilon }\gamma ^{d}\psi _{d}\right) -2\widetilde{%
R_{\mu }^{1}\,^{a}}\left( \overline{\varepsilon }\widetilde{\gamma }^{\mu
}\psi _{a}\right) \right] \right\} +H.C.C.\,.
\end{eqnarray}
Noting that the two terms in the curly brackets are the H.C.C. of one
another, we finally get : 
\begin{equation}
\delta \mathcal{L}_{\psi }^{0^{\prime }}=-\imath k\left( e\widetilde{e}%
\right) ^{\frac{1}{2}}\left[ R^{1}\left( \overline{\varepsilon }\gamma
^{d}\psi _{d}\right) -2R_{\mu }^{1}\,^{a}\left( \overline{\varepsilon }%
\gamma ^{\mu }\psi _{a}\right) \right] +H.C.C.\,.
\end{equation}
This term effectively cancels (50) if the constant $k$ is chosen to be $k=-%
\frac{\imath }{2}$.

Now the expression (53) is identical to (56), and then cancels (50), if and
only if the two following identifications are simultaneously satisfied : 
\begin{eqnarray}
\Omega ^{\mu \nu \alpha \rho } &=&\left( e\widetilde{e}\right) ^{\frac{1}{2}%
}\left( \eta ^{\nu \rho \alpha \widetilde{\mu }}+\eta ^{\nu \rho \widetilde{%
\alpha }\mu }\right) \,, \\
\Omega ^{\mu \nu \alpha \rho } &=&\left( e\widetilde{e}\right) ^{\frac{1}{2}%
}\left( \eta ^{\mu \nu \alpha \widetilde{\rho }}+\eta ^{\mu \nu \widetilde{%
\alpha }\rho }\right) \,.
\end{eqnarray}
However the two expressions are not compatible as one can readily verify.
Indeed, chosing one of the preceding expressions, we inevitably end up with
either an $\eta $ term that does effectively have the right antisymmetry
properties but gives rise to contractions of $R_{\mu \nu }^{1}\,^{ab}$ with
two vierbeins ($e_{\bullet }^{\alpha }e_{\bullet }^{\rho }$), whereas a
vierbein and a conjugate vierbein are required ($e_{\bullet }^{\mu }%
\widetilde{e}_{\bullet }^{\nu }$), or with an $\eta $ term that does not
have the right antisymmetry property that can lead to $R_{\mu \nu
}^{1}\,^{ab}$, via the commutator of two covariant derivatives. Thus, there
is no linear combination $\Omega ^{\mu \nu \alpha \rho }$ that can give
rise, through the variation (51), to the linear variation in $\psi _{\mu }$
that might cancel (50). It is worth emphasizing, once more, that neither the
variation of $\mathcal{L}_{NGT}^{2}$ (38) nor that of $\mathcal{L}_{\chi }$
and $\mathcal{L}_{\varphi }$ (42)-(43) can give rise to contributions to the
cancellation of (50).

One may wonder if the massive version of NGT would make things work better.
This is not the case because i) the massless and the massive versions of NGT
have the same ''kinetic'' part (4) (or (31)) in their Lagrangians and ii)
both versions describe a massless graviton, to which one associates a
massless gravitino, and thus one can adopt the generalization of the
Rarita-Schwinger Lagrangian (39). We thus end up with the same problem(s) as
with the massless version.

Finally, we note that the undesired extra-terms generated by (66) or (67) in
the variation $\delta \mathcal{L}_{\psi }$ linear in $\psi _{\mu }$ suggest 
\emph{non-minimal} models in which the linear variation in $\psi _{\mu }$
vanishes, but at the cost of adding terms to the Lagrangian (36) and
slightly modifying the infinitesimal supersymmetric law of transformation
(51). One example, among others, is given by the following choices : 
\begin{eqnarray}
\Omega ^{\mu \nu \alpha \rho } &=&\left( e\widetilde{e}\right) ^{\frac{1}{2}%
}\left( \eta ^{\mu \nu \alpha \rho }+\eta ^{\mu \nu \widetilde{\alpha }\rho
}\right)  \nonumber \\
&=&\left( e\widetilde{e}\right) ^{\frac{1}{2}}\eta ^{abcd}\left( e_{a}^{\mu
}e_{b}^{\nu }e_{c}^{\alpha }e_{d}^{\rho }+e_{a}^{\mu }e_{b}^{\nu }\widetilde{%
e}_{c}^{\alpha }e_{d}^{\rho }\right) ,
\end{eqnarray}
and 
\begin{equation}
\delta _{0}\psi _{\mu }=\mathcal{D}_{\mu }\varepsilon +e_{\mu }^{a}%
\widetilde{e}_{a}^{\beta }\mathcal{D}_{\beta }\varepsilon ,
\end{equation}
along with the addition of the following term to the NGT Lagrangian (36) : 
\begin{equation}
\mathcal{L}_{NGT}^{\prime }=-\frac{1}{4}\left( e\widetilde{e}\right) ^{\frac{%
1}{2}}e_{a}^{\mu }e_{b}^{\nu }R_{\mu \nu }^{1}\,^{ab}+H.C.C.\,.
\end{equation}
We have not pursued the investigation of the full invariance of such a model
because it does not seam reliable ; its drawback, to say the least, is that
the interpretation, in the linear approximation, in terms of particles is
seriously spoiled by the addition of extra terms like (70).

\section{Conclusion}

In this paper, we have tried to build a locally supersymmetric minimal model
based on nonsymmetric gravitation theory. We have adopted the most
straightforward generalizations from ordinary supergravity, to the case of
NGT. This trial led us to a no-go result, namely the impossibility of
cancelling the terms linear in the gravitino field, coming from the
variation of the graviton and gravitino Lagrangians. It is worth stressing
again that this impossibility stands equally well for both the massless and
the massive versions of NGT, even though we have not considered variations
of non geometrical terms, since these cannot contibute to that of the
kinetic part of the lagrangian which is common to both versions. This no-go
result relies first on the choices we made in our model, and second on the
hyperbolic complex structure introduced to accomodate the vierbein formalism
of NGT. Nevertheless, it points out the serious difficulties facing the
construction of a supersymmetric theory of NGT.

Another alternative is to use the supersymmetric extension of the full local
invariance group $GL(4,{R)}\sim U(3,1,H)$ of NGT, instead of the
superPoincar\'{e} group, and then to look for some symmetry breaking
mechanism to reduce the former to the latter. This would entail the use of
infinite component fields \cite{20}. In addition to the mathematical
difficulties that one would face in such an approach, its physical
interpretation is not obvious.

Finally, we note that beyond the mathematical difficulties raised by the
introduction of the nonsymmetric metric tensor $g_{\mu \nu }$, lies the lack
of physical interpretation of the antisymmetric part of this tensor, which
seem to be at the heart of the problem(s) confronting NGT.

\newpage

\noindent \textbf{{\large {Acknowledgement}}}

I am very grateful to G\'erard Cl\'{e}ment for his support and helpful
private communications.


\bigskip

\bigskip

\begin{thebibliography}{99}
\bibitem{1}  J.\ W.\ Moffat, \emph{Phys. Rev. }\textbf{D19}, (1979) 3554.

\bibitem{2}  A. Einstein, \emph{Ann. Math.} \textbf{46,} (1945) 578.

A.\ Einstein and E.G.\ Strauss, \emph{Ann. Math.} \textbf{47,} (1946) 731.

\bibitem{3}  G. Kunstatter, J.W.\ Moffat and P.\ Savaria, \emph{Phys. Rev. }%
\textbf{D19}, (1979) 3559.

J.\ W.\ Moffat, \emph{Phys. Rev. }\textbf{D19}, (1979) 3562.

J.\ W.\ Moffat, \emph{Phys. Rev. }\textbf{D35}, (1987) 3733 ; Erratum-ibid. 
\textbf{D36}, (1987) 3290.

J. W. Moffat and E.\ Woolgar, \emph{Phys. Rev; }\textbf{D37} (1988) 918.

\bibitem{4}  G.\ Kunstatter, J.W.\ Moffat and J.\ Malzan, \emph{J.\ Math.
Phys. }\textbf{24}, (1983) 886.

\bibitem{5}  Z.Z. Zhong, \emph{J.\ Math. Phys. }\textbf{25}, (1984) 3538.

Z.Z. Zhong, \emph{J.\ Math. Phys. }\textbf{26}, (1985) 404.

\bibitem{6}  J.W.\ Moffat, \emph{J. Math. Phys. }\textbf{29}, (1988) 1655.

\bibitem{7}  R.B.\ Mann and J.W.\ Moffat, \emph{J.\ Phys. A }\textbf{14},
(1981) 2367.

\bibitem{8}  T.\ Damour, S.\ Desesr and J. McCarthy, \emph{Phys. Rev. }%
\textbf{D45, }(1992) R3289.

T.\ Damour, S.\ Desesr and J. McCarthy, \emph{Phys. Rev. }\textbf{D47, }%
(1993) 1541.

\bibitem{9}  X.\ Bekaert, B.\ Knaepen and C.\ Schomblond, \emph{Phys. Lett. }%
\textbf{B481}, (2000) 89.

\bibitem{10}  N.J.\ Cornish and J.W.\ Moffat, \emph{Phys. Rev. }\textbf{D47, 
}(1993) 4421.

N.J.\ Cornish, J.W.\ Moffat and D.C. Tatarski,\emph{\ Phys. Lett. A }\textbf{%
173,} (1993) 109.

N.J. Cornish, J.W. Moffat and D.C. Tatarski, \emph{Gen. Rel. Grav. }\textbf{%
27}, (1995) 933.

J.\ W.\ Moffat, University of Toronto Preprint \textbf{UTPT-93-11}, {%
gr-qc/9306003.}

\bibitem{11}  J.W.\ Moffat, \emph{Phys. Lett. B }\textbf{355, }(1995) 447.

J.\ L\'{e}gar\'{e} and J.W.\ Moffat, \emph{Gen. Rel. Grav. }\textbf{27, }%
(1995) 761.

\bibitem{12}  J.W.\ Moffat, \emph{J.\ Math. Phys. }\textbf{36}, (1995) 3722.

\bibitem{13}  D.\ Freedman, S.\ Ferrara and P. van Nieuwenhuizen, \emph{%
Phys. Rev.} \textbf{D13}, (1976) 3214.

S.\ Deser and B.\ Zumino, \emph{Phys. Lett.} \textbf{62B}, (1976) 335.

\bibitem{14}  A.\ H.\ Chamseddine and P.\ C.\ West, \emph{Nucl. Phys. }%
\textbf{B129}, (1977) 39.

\bibitem{15}  Y.\ Ne'eman, \emph{Ann. Inst. H.\ Poincar\'{e}} \textbf{A 28, }%
369 (1978).

\bibitem{16}  M.A.\ Tonnelat, ''\emph{La Th\'{e}orie du Champ Unifi\'{e}
d'Einstein}'' (Gauthier-Villars, Paris, 1955).

\bibitem{17}  K. Ait Moussa and N. Mebarki, \emph{Acta Phys. Pol. }\textbf{%
B23}, (1992) 1195.

\bibitem{18}  F.\ Khelili, J.\ Mimouni and N.\ Mebarki, \emph{J. Math. Phys. 
}\textbf{42}, (2001) 3615.

\bibitem{19}  P.\ van Nieuwenhuisen, \emph{Phys. Rep.} \textbf{68 N}$%
{{}^\circ}%
$\textbf{4}, (1981) 189.

\bibitem{20}  J.\ W.\ Moffat, \emph{Phys. Lett. B} \textbf{206}, (1988) 499.
\end{thebibliography}
\end{document}